\documentclass[aps,pre,preprint,superscriptaddress]{revtex4-1}

\usepackage[T1]{fontenc}       
\usepackage{graphicx}
\usepackage{dcolumn}
\usepackage{bm}
\usepackage{amsmath}
\usepackage{amsfonts}
\usepackage{amssymb}
\usepackage{braket}
\usepackage[separate-uncertainty = true,multi-part-units=single]{siunitx}
\usepackage{hyperref}
\usepackage{subcaption}
\usepackage{color}
\usepackage{tabularx}

\let\oldsqrt\sqrt
\def\sqrt{\mathpalette\DHLhksqrt}
\def\DHLhksqrt#1#2{%
\setbox0=\hbox{$#1\oldsqrt{#2\,}$}\dimen0=\ht0
\advance\dimen0-0.2\ht0
\setbox2=\hbox{\vrule height\ht0 depth -\dimen0}%
{\box0\lower0.4pt\box2}}

\begin{document}
\preprint{ET11583}
\title{Shear deformation of low-density polymer brushes in a good solvent}

\author{Airidas Korolkovas}
\affiliation{Institut Laue-Langevin, 71 rue des Martyrs, 38000 Grenoble, France}
\affiliation{Department for Physics and Astronomy, L{\"a}gerhyddsv{\"a}gen 1, 752 37 Uppsala, Sweden}
\email{korolkovas@ill.fr}
\author{Philipp Gutfreund}
\affiliation{Institut Laue-Langevin, 71 rue des Martyrs, 38000 Grenoble, France}
\author{Alexis Chennevi{\`e}re}
\affiliation{Laboratoire de physique des solides, CNRS, Universit{\'e} Paris-Sud, 91405 Orsay, France}
\author{John F. Ankner}
\affiliation{Spallation Neutron Source, Oak Ridge National Laboratory, 37830 Oak Ridge, TN, USA}
\author{Franz A. Adlmann}
\affiliation{Department for Physics and Astronomy, L{\"a}gerhyddsv{\"a}gen 1, 752 37 Uppsala, Sweden}
\author{Max Wolff}
\affiliation{Department for Physics and Astronomy, L{\"a}gerhyddsv{\"a}gen 1, 752 37 Uppsala, Sweden}
\author{Jean-Louis Barrat}
\affiliation{Universit\'{e} Grenoble Alpes, LiPhy, 38402 Saint Martin d'H\`{e}res, France}

\date{\today}

\begin{abstract}
Self-consistent field approach is used to model a single end-tethered polymer chain on a substrate subject to various forces in three dimensions. Starting from a continuous Gaussian chain model, the following perturbations are considered: (i) hydrodynamic interaction with an externally imposed shear flow for which a new theoretical framework is formulated; (ii) excluded volume effect in a good solvent, treated in a mean field approximation; (iii) monomer-substrate repulsion. While the chain stretches along the flow, the change of the density profile perpendicular to the substrate is negligible for any reasonable simulation parameters. This null effect is in agreement with multiple neutron scattering studies.
\end{abstract}

\maketitle
\section{Introduction}
End-tethered macromolecules have attracted much interest from pure academic research~\cite{Alexander1977} to applications in biology and material science~\cite{Azzaroni2012}. The equilibrium structure of a polymer brush in contact with a polymer melt~\cite{Clarke1995}, a solvent, or a polymer solution~\cite{Jones1999,Currie2003}, has been quite well understood in theory, first using scaling arguments \cite{Alexander1977,deGennes1980} and later extended quantitatively by mean-field calculations~\cite{Milner1991}. The theoretical results were confirmed by molecular dynamics (MD) simulations~\cite{Grest1999,Hoy2007}. Early experiments have suffered from a limited availability of grafting densities and chain lengths, but eventually could corroborate the predicted scaling laws~\cite{Zhao2000}. The density profile of the brush was revealed by small angle neutron scattering (SANS) and neutron reflectometry (NR)~\cite{Grest1999}.

Investigations of polymer brushes under shear are becoming commonplace since many of their potential applications invoke a shear force, as detailed in a recent review~\cite{kreer2016polymer}. A large body of theoretical work was done using de Gennes model for weakly grafted films~\cite{Brochard1992,BrochardWyart1996} and for high grafting densities~\cite{Rabin1990,barrat1992,Gay1999}, in addition to MD simulations on high density films~\cite{Grest1999,pastorino2006,muller2008}. All theories agree that grafted chains stretch along the flow when under a sufficiently strong shear rate. The lateral stretch has been measured experimentally with atomic force microscopy (AFM)~\cite{kato2003polymer}. However, AFM is an invasive probe and requires an application of a normal force to operate, provoking a decrease of the brush height, linearly proportional to the normal force. In opposition, an increase of the brush height has been indirectly inferred using surface force apparatus (SFA)~\cite{klein1991forces}. This effect was later explained theoretically by thermodynamic arguments~\cite{barrat1992}. 

Neutron scattering, while advantageous as a direct and a non-invasive probe, bears certain limitations. First, low density brushes eventually become too faint to detect, unlike in AFM where a single mushroom can be imaged. Second, the lateral brush structure has never been reported, although it is hypothetically possible to deuterate a fraction of the chains and measure their form factor using rheo-grazing incidence SANS~\cite{wolff2008shear, newby2015situ}. On the flip side, neutrons excel at measuring the density profile perpendicular to the substrate. While the effect of shear on this axis is only secondary and hence weaker, recent NR studies on melts~\cite{Sasa2011, chenneviere2016direct} and semi-dilute solutions~\cite{korolkovas2017polymer}, both well entangled with the brushes, have demonstrated up to \SI{20}{\percent} decrease of brush thickness perpendicular to flow. This decrease of thickness is a quadratic function of the shear rate, in contrast to the decrease seen in non-entangled AFM studies, which is a linear function of the normal force~\cite{kato2003polymer}.

The shrinkage of entangled brushes is attributed to the normal stress difference of the bulk fluid, and is fundamentally the same effect as for entangled bulk chains~\cite{muller1993chain, korolkovas2018anisotropy}. This phenomenon does not occur in non-entangled, Newtonian liquids, as evidenced by a SANS study of bulk flow containing long but dilute polystyrene chains, found to stretch by \SI{50}{\percent} along flow, and no change perpendicular to flow~\cite{lindner1988shear}. The same conclusion is drawn for grafted chains, where multiple NR studies have reported a null effect~\cite{baker2000,ivkov2001}. These experiments have used rather high density brushes, which screen the flow, possibly reducing the interaction with the solvent. In the present article we continue the experimental search for an effect, using a lower grafting density to increase the solvent penetration. Further, we use longer chains which have a longer relaxation time and couple stronger to the experimentally accessible shear rates. 

The theory of low density mushroom brushes is less developed and the brush response to shear flow is currently unknown. To fill this gap, we present a novel mean-field algorithm, which considers the full three-dimensional density field of an isolated mushroom, interacting self-consistently with the velocity field of the sheared solvent. The excluded volume and the monomer-substrate repulsions are added as well.

\section{Experimental}
\begin{figure}[ptbh!] 
\begin{subfigure}{.45\textwidth}
		\includegraphics[width=\linewidth]{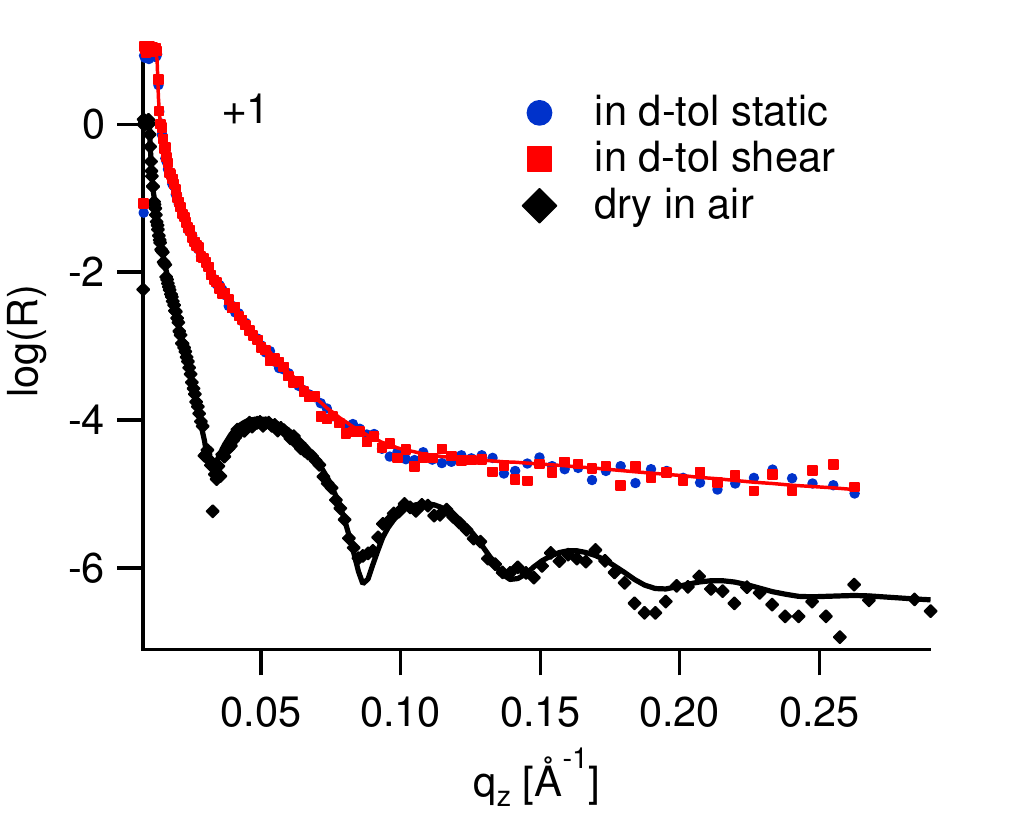}
\end{subfigure}
\hfill
\begin{subfigure}{.53\textwidth}
		\includegraphics[width=\linewidth]{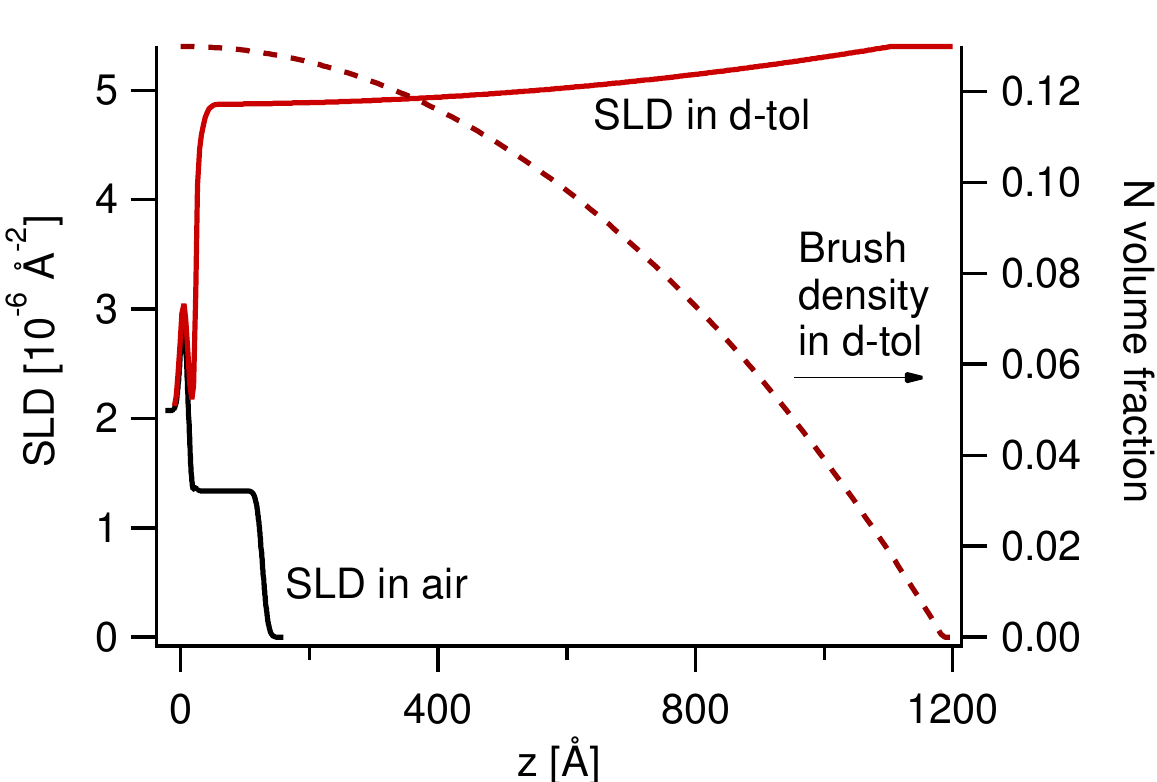}
\end{subfigure}
\caption{Left panel: experimental reflectivity data on a logarithmic scale for a protonated PS brush $M_n=\SI{250}{\kilo\gram\per\mole}$ in air (black diamonds), in d-toluene static (blue circles) and in d-toluene at a shear rate of \SI{500}{\per\second} (red squares). Right panel: fitted SLD profiles, where the dashed line is the corresponding monomer density of the brush, shown on the right axis.}\label{NR}
\end{figure}

Amino end-functionalized polystyrene (PS) was synthesized in-house to a molecular weight of $M_n=\SI{250}{\kilo\gram\per\mole}$ ($N=2400$) and a polydispersity of 1.4. It was grafted onto a self-assembled monolayer (SAM) of diethoxy(3-glycidyloxypropyl)methysilane) (\SI{97}{\percent} Sigma) deposited on a \SI[product-units=power]{70 x 70 x 10}{mm} single crystal silicon block (100, Crystec, Berlin). Details about the sample preparation can be found in Ref.~\cite{chenneviere2013quantitative}. NR in air has revealed a silicon oxide thickness of \SI{1.05}{\nano\meter} and a brush thickness of \SI{10.7}{\nano\meter} with a scattering length density (SLD) of \SI{1.34e-6}{\per\square\angstrom}. Next, the brush was put into contact with deuterated toluene-d$_{8}$ (Sigma-Aldrich, \SI{99.6}{\percent} deuteration) at a temperature of \SI{18.5\pm 1.5}{\celsius}. The resulting NR and the corresponding SLD are shown in Fig.~\ref{NR}. The reflectivity was fitted with a parabolic density profile~\cite{zhulina1991theory}, revealing a swollen thickness of \SI{120}{\nano\meter}.

\begin{table}[htb]
\begin{tabularx}{0.6\textwidth}{X X X X X}
Molecular weight, kg/mol &
Dry height, nm &
Wet height, nm &
Shear rate, 1/s &
Ref.\\
\hline
83 & 17.5 & 75 & 130000 & \cite{ivkov2001}\\
184 & -- & 80 & 10000 & \cite{baker2000}\\
250 & 10.7 & 120 & 500 & our data\\
280 & -- & -- & 8500 & \cite{lindner1988shear} (bulk)
\end{tabularx}\caption{PS brushes in toluene under shear reported in various NR studies. Bulk rheo-SANS~\cite{lindner1988shear} is included for context. None of these experiments could detect any density change perpendicular to flow.}\label{explist}
\end{table}

A steady shear flow of \SI{500}{\per\second} was applied using \textit{Anton Paar MCR 501} rheometer in a previously described setup~\cite{wolff2013combined}. The resulting NR profile is shown in Fig.~\ref{NR} and is seen to be unaffected by the shear. Our experimental conditions are summarized together with literature data in Table~\ref{explist}.

\section{Theory}

\begin{figure}[htb]
    \centering
    \includegraphics[width=0.8\linewidth]{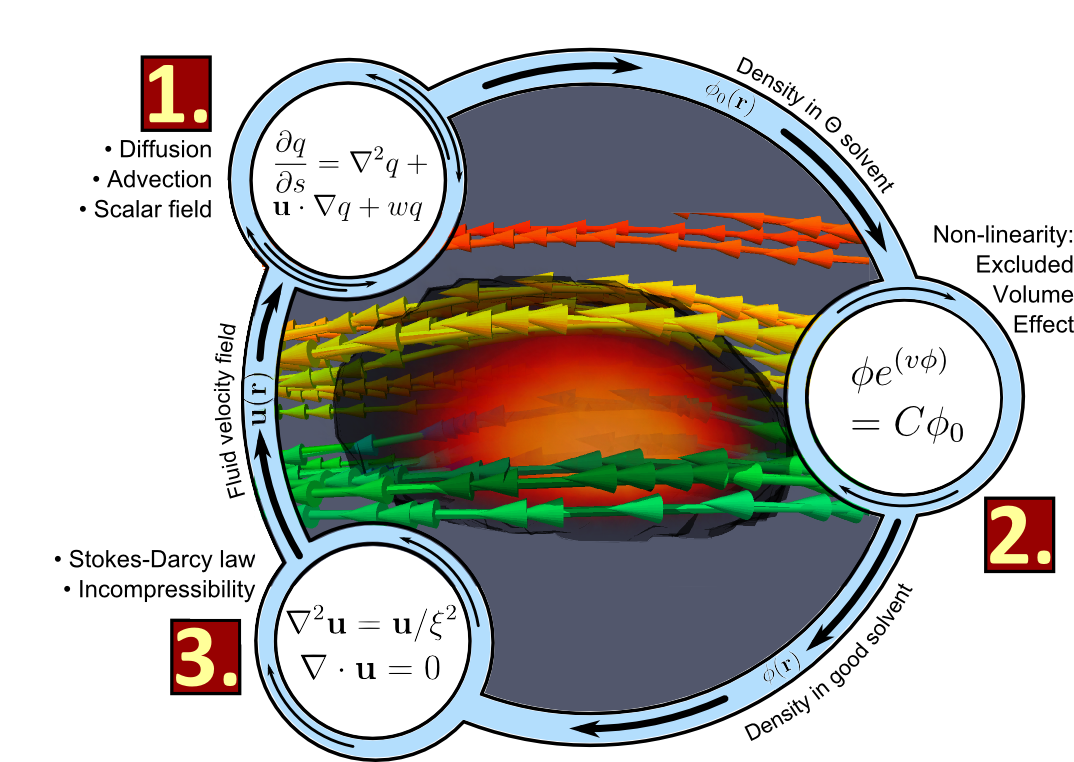} 
    \caption{Simulation summary. The semi-transparent cloud in the center shows the mushroom density profile (orange is denser). The streamlines of the solvent flow are visualized with cone glyphs, seeded at three different heights: green near the substrate, yellow in the middle of the brush, red at the top. The velocity field deviates from laminar, flowing around the core of the mushroom. As a result, there is upwards drag on the right side of the mushroom but also an equal and opposite downwards pull on the left side. The net vertical force is close to zero, resulting in zero change of the density profile perpendicular to the substrate. The simplified equations around the picture show key steps of the algorithm, described in detail in the main text.}\label{algorithm}
\end{figure}

Here we propose a field-based mathematical model of an isolated polymer chain end-tethered to a substrate and subject to shear flow, as well as excluded volume effect and a monomer-substrate repulsive force. All of these forces are accounted for simultaneously, providing the polymer density distribution consistent with the distorted velocity field of the sheared solvent. The general framework is that of self consistent field theory~\cite{fredrickson2006}, with nontrivial modifications needed for a correct description of an external flow field. An outline of the simulation algorithm and a snapshot of the main result is shown in Fig.~\ref{algorithm}.

To start off, we model our isolated polymer chain as a continuous Gaussian coil, meaning that the the $s$-th monomer is labeled by a continuous variable $s \in [0,1]$. The partition function $q$ and its conjugate $q^{\dagger}$ of such a chain in an external potential $w(\mathbf{r})$ obey the modified diffusion equation (MDE)~\cite{edwards1965,degennes1969}:
\begin{align}
\label{mde1} \frac{\partial q}{\partial s} &= R_g^2 \nabla^2 q - \frac{w(\mathbf{r})}{k_B T}q,\quad q(\mathbf{r},s=0) = \delta(\mathbf{r}),\\ 
\label{mde2} \frac{\partial q^\dagger}{\partial s} &= - \left(R_g^2 \nabla^2 q^\dagger - \frac{w(\mathbf{r})}{k_B T}q^\dagger\right), \quad q^\dagger(\mathbf{r},s=1) = 1. 
\end{align}
The radius of gyration $R_g = \sqrt{Na^2/6}$ defines the natural unit of length, and refers to the size of an ideal random walk of $N$ steps of length $a$. The solution $q(\mathbf{r},s)$ is the partition function for a polymer chain of length $s$ starting at $\mathbf{r_0} = 0$ (the tethering point) and ending at an arbitrary point $\mathbf{r}$. Likewise, $q^\dagger (\mathbf{r},s)$ is the partition function for a polymer of length $(1-s)$ having a uniform distribution at its $s=1$ end, and terminating at $\mathbf{r}$ as explained in more detail in Ref.~\citenum{matsen2006}. The total partition function is obtained by summing over all the intermediate positions $\mathbf{r}$:
\begin{equation}
Q[w] = \int d\mathbf{r}\, q(\mathbf{r},s) q^\dagger(\mathbf{r},s).
\end{equation}
One can check using integration by parts that $Q$ is independent of the monomer $s$ at which it is evaluated: $dQ/ds \equiv 0$. The practical result of the MDE is the polymer density, obtained by summing the contribution from each segment $s$ and normalizing by the $Q$:
\begin{equation}\label{rhogen}
\rho(\mathbf{r}) = \frac{1}{Q} \int_0^1 ds\, q(\mathbf{r},s) q^\dagger(\mathbf{r},s).
\end{equation}
It is straightforward to verify that this density is normalized: $\int d\mathbf{r}\, \rho(\mathbf{r}) \equiv 1$.

The simplest application of the theory stated so far is the ideal Gaussian chain which is purely governed by the maximization of entropy. In this case the potential energy $w(\mathbf{r}) = 0$, and the solution to Eqs.~\eqref{mde1}-\eqref{mde2} is $q(\mathbf{r},s) = (4\pi R_g^2 s)^{-3/2}e^{-r^2/4sR_g^2}$; $q^\dagger (\mathbf{r},s) = 1$. The total partition function $Q=1$, while the density is 
\begin{equation}\label{rho}
\rho(\mathbf{r}) = \int_0^1 ds\, \frac{1}{(4\pi R_g^2 s)^{3/2}} \exp\left(-\frac{r^2}{4R_g^2 s}\right).
\end{equation}

We will now proceed to compute how the polymer density distribution $\rho(\mathbf{r})$ changes under an external shear flow, as well as excluded volume and surface repulsion.

\subsection{Applied shear flow}
Consider that the chain is placed in an external solvent flow described by the velocity field  $\mathbf{u}(\mathbf{r}) $, which for the moment will be assumed to be fixed and insensitive to the chain conformation. An example is a linear shear flow $\mathbf{u}(\mathbf{r}) = \dot{\gamma}z\mathbf{\hat{x}}$. This will exert a Stokes force $\mathbf{F} = \zeta \mathbf{u}$ on the monomers, where $\zeta=6\pi \eta a$ is the monomeric friction coefficient, with $a$ the monomer size and $\eta$ the solvent viscosity. Unfortunately, such a force cannot be derived from a scalar potential because its curl is non-zero: $\nabla \times \mathbf{F} \neq 0$, and hence the energy of the polymer chain is ill-defined, as it depends on the path taken by the chain (for a related discussion, see Appendix).

Our main novelty to handle this non-conservative aspect of the hydrodynamic forces is to first obtain the elementary propagator for a small chain segment, and then use the Kolmogorov-Chapman equation to derive the full partition function. Within the length scale of one bond length we can safely consider the speed $\mathbf{u}$ to be uniform,  in which case the propagator for an elementary chain segment of length $\epsilon$ stretching between $\mathbf{r}$ and $\mathbf{r} +\Delta \mathbf{r}$ remains Gaussian, with a bias due to the uniform velocity field:
\begin{equation}
P(\Delta \mathbf{r},\mathbf{r},\epsilon) = \left(\frac{k}{2\pi k_B T}\right)^{3/2} \exp \left[-\frac{k(\Delta \mathbf{r}-\mathbf{u}(\mathbf{r})\zeta/k)^2}{2k_B T}\right].
\end{equation}
where
 $k=k_B T/(2R_g^2 \epsilon)$ is the stiffness of the segment. Inserting this into the Kolmogorov-Chapman equation 
\begin{equation}
q(\mathbf{r},\mathbf{r_0},s+\epsilon) = \int d(\Delta \mathbf{r})  q(\mathbf{r}-\Delta \mathbf{r},\mathbf{r_0},s+\epsilon)  P(\Delta\mathbf{r},\mathbf{r}-\Delta\mathbf{r},\epsilon)
\end{equation}
we obtain a diffusion-advection kind of equation for the complete partition function:
\begin{equation}\label{adv1}
\frac{\partial q}{\partial s} = R_g^2 \nabla^2 q - \tau_r \mathbf{u}\cdot \nabla q, 
\end{equation}
where $\tau_r = 2\zeta N R_g^2 /(k_B T)$ is the Rouse relaxation time of the polymer~\cite{doi1986}. A similar reasoning for the complementary partition function yields
\begin{equation}\label{adv2}
\frac{\partial q^\dagger}{\partial s} = -R_g^2 \nabla^2 q^\dagger - \tau_r \mathbf{u}\cdot \nabla q^\dagger.
\end{equation}
Once again we verify that the total partition function $Q = \int d\mathbf{r}\, qq^\dagger$ is independent of $s$.
\begin{figure}[htb]
    \centering
		\includegraphics[width=0.8\linewidth]{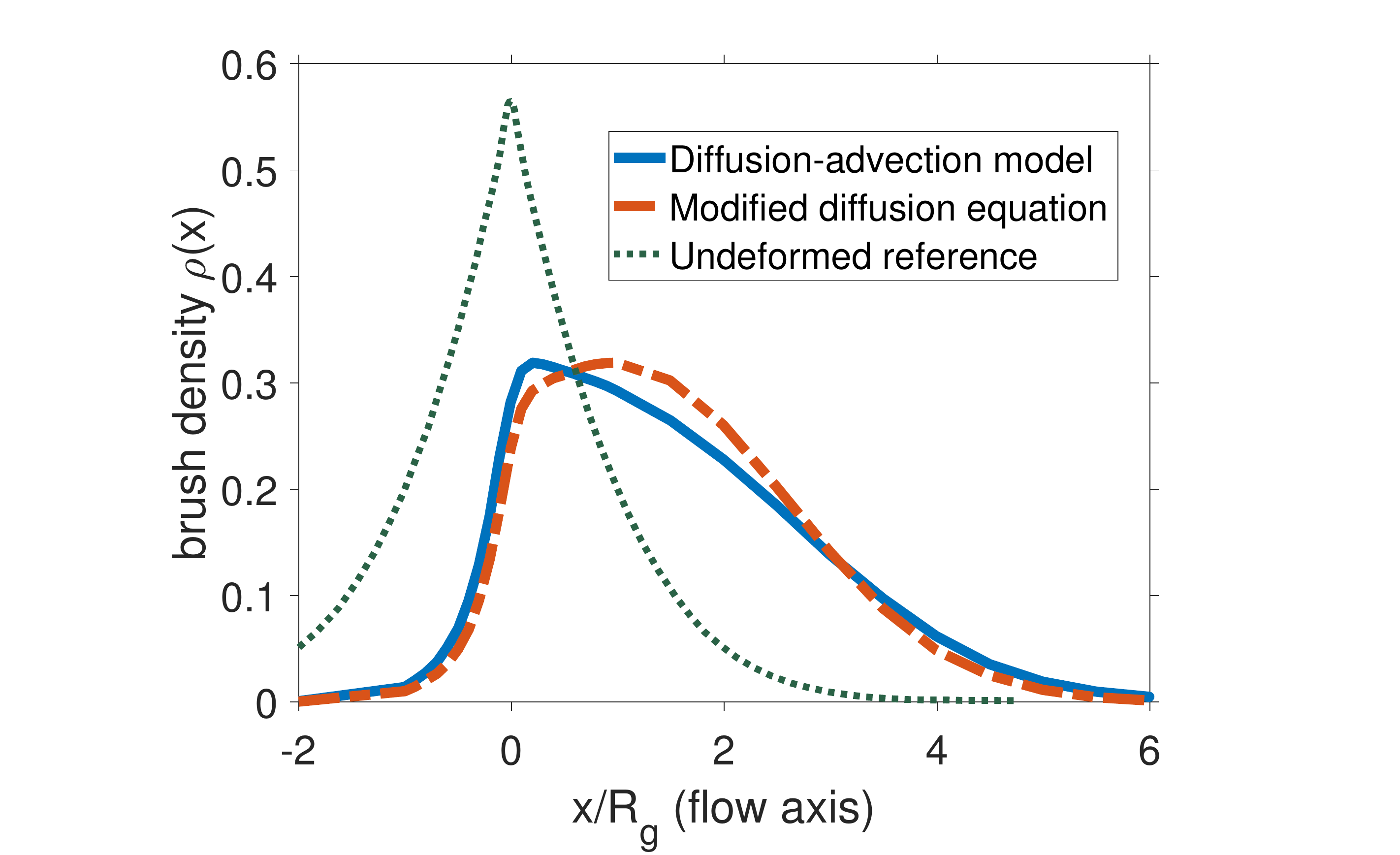}
    \caption{Comparison between the modified diffusion equation and the diffusion-advection equation, both models of a uniform flow field. Parameters used in this example are $u=3$, $g=9/4$ (see Appendix).}\label{uflow}
\end{figure}
To test our equation, we consider the case of uniform flow $\mathbf{u} = \text{const.}$ for which an analytical solution is derived (see Appendix). Earlier works (i.e. Ref.~\cite{lai1993}) have modeled such uniform flow by the regular MDE, Eqs.~\eqref{mde1}-\eqref{mde2}, using a scalar potential $w = -\zeta u x$. The two solutions are compared in Fig.~\ref{uflow}, showing the difference between our diffusion-advection theory and the scalar theory.

Not only the fluid velocity $\mathbf{u(r)}$ exerts a force on the polymer, but also the presence of the polymer disturbs the flow. To take this effect into account, we assume that the polymer can be treated as a porous medium whose resistivity is proportional to the monomer density $\rho(\mathbf{r})$ (different dependencies could also be used but they typically bring only small modifications to the final result). The flow profile is obtained from the stationary Stokes-Darcy law for incompressible flow, using the density obtained from Eqs.~\eqref{adv1}-\eqref{adv2}:
\begin{equation}\label{darcy}
\nabla^2 \mathbf{u} - \frac{1}{\eta}\nabla p = \frac{\rho(\mathbf{r})}{\xi^2}\mathbf{u} \quad \text{and} \quad \nabla \cdot \mathbf{u} = 0,
\end{equation}
where $\xi$ is a constant hydrodynamic screening length. We use a Dirichlet boundary condition with $\mathbf{u} = \dot{\gamma} z\mathbf{\hat{x}}$ far away from the brush. (Note: This set of four coupled PDEs can be solved by adding a penalty term $\lambda p$ to the divergence relationship, with $\lambda \rightarrow 0$. In numerical schemes beware that the pressure and the velocity fields must be defined on separate LBB-compatible finite element spaces~\cite{brezzi1974}). The solution $\mathbf{u}$ is fed back to Eqs.~\eqref{adv1}-\eqref{adv2} and the process is repeated until convergence is achieved (which usually happens within 4-7 iterations).

\subsection{Excluded volume effect}
At the mean field level, the excluded volume effect is achieved through a repulsive potential $w(\mathbf{r}) = vN\rho(\mathbf{r}) k_B T$ where $v$ is the excluded volume per monomer and $k_B T$ is the thermal energy. This approach was pioneered by Edwards, who reproduced the Flory scaling $R_F \propto aN^{3/5}$, as expected from the mean field character. In the present study we need to take into account the excluded volume effect for chain configurations which are distorted from a spherical symmetry and hence a numerical approach is required to obtain the density profiles.

If we plug in the Flory potential $vN\rho(\mathbf{r})$ into the MDE, the result is a set of two coupled partial non-linear integro-differential equations with both the non-linearity and the coupling on the integral term. To solve it, one may attempt a Picard iteration: start with a Gaussian coil Eq.~\eqref{rho}, plug in the density to the MDE, solve for the new density with excluded volume, plug in the density, and repeat until convergence. Numerically, this scheme only converges when the energy of the excluded volume interaction is small compared to the thermal energy: $vN^2/R_F^3 \approx (v/a)^3 N^{1/5}\ll 1$, which does not apply for long chains $N\gg 1$ in a good solvent $v\approx a^3$.

To target realistic conditions, we propose the following algorithm for rapid and stable convergence. First, we replace the partition functions by
\begin{equation}
q(\mathbf{r},s) \, \rightarrow \, q_0(\mathbf{r},s) \exp(-svN\rho) \quad \text{and} \quad q^\dagger (\mathbf{r},s) \, \rightarrow \, q_0^\dagger (\mathbf{r},s) \exp\left[-(1-s)vN\rho\right].
\end{equation}
This eliminates the problematic Flory term, at the expense of adding derivatives of $\rho$. However, we can neglect these derivatives, based on the following argument. Knowing that the final density profile will be some monotonically decaying function like $\rho \approx \exp(-z/R_F)$, we can estimate the magnitude of the second derivative as: $(R_g/R_F)^2 \exp(-z/R_F) \propto N^{-1/5}$. This has to be contrasted with the second derivative of $q_0$, which decays like $R_g^2 q_0/R_g^2 \propto 1$. Hence, for a very long chain $N\gg 1$, the extra derivative of $\rho$ is negligible.

In the second step, we solve the MDE without the Flory term to obtain the partition functions $q_0^\dagger (\mathbf{r},s)$ and $q_0^\dagger (\mathbf{r},s)$. The density is given by Eq.~\eqref{rhogen} as usual:
\begin{equation}\label{exvol}
\rho = C \int_0^1 ds\, qq^\dagger = C\int_0^1 ds\, q_0 {q_0}^\dagger \exp\left(-vN\rho\right) = C\rho_0 \exp\left(-vN\rho\right).
\end{equation}

In the third step, we iteratively solve Eq.~\eqref{exvol} to obtain the function $\rho(\mathbf{r})$ and the constant $C$ subject to normalization constraint $\int d\mathbf{r}\, \rho = 1$. We have to take its logarithm in order to damp the errors at each iteration, as opposed to amplifying them with the exponent: $vN\rho = \ln C + \ln (\rho_0/\rho)$. Let us assume that the first guess function is given by $\rho = \rho_1$, where $\rho_1$ is normalized. An improvement would be $\rho_2 = \rho_1 + \Delta \rho_1$, where $\Delta \rho_1$ is small. We now Taylor-expand our equation to obtain a correction
\begin{equation}\label{drho}
\Delta \rho_1 = \rho_1 \left[\frac{\ln C_2 + \ln (\rho_0/\rho_1) -vN\rho_1}{1+vN\rho_1}\right].
\end{equation}
The insofar unknown constant $C_2$ is fixed by requiring $\int d\mathbf{r}\, \Delta \rho_1 = 0$. A good initial guess is simply $\rho_1 = \rho_0$, i.e. the density in theta solvent. A mere 5-6 iterations usually suffice to reach convergence of Eq.~\eqref{drho}. We observe a redistribution of polymer from the core to the periphery, thus leveling off the density as anticipated. A more uniform density is also what justifies the approximation we made by neglecting derivatives of $\rho$ in the MDE.

We have applied this algorithm to estimate the density profile of a polystyrene chain in toluene (a good solvent) when the chain is end-tethered to a substrate. The results are shown in Fig.~\ref{slice} and the related discussion is in the Appendix.

\begin{figure}[tbph]
  \begin{subfigure}[b]{0.5\linewidth}
    \centering
    \includegraphics[width=\linewidth]{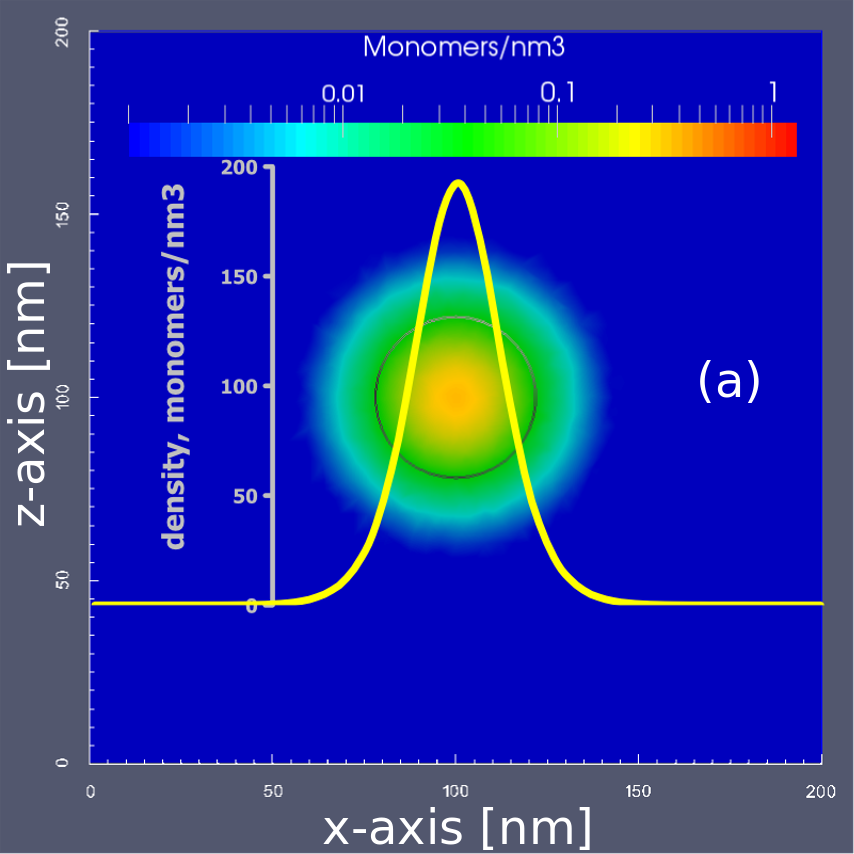} 
  \end{subfigure}
  \begin{subfigure}[b]{0.5\linewidth}
    \centering
    \includegraphics[width=\linewidth]{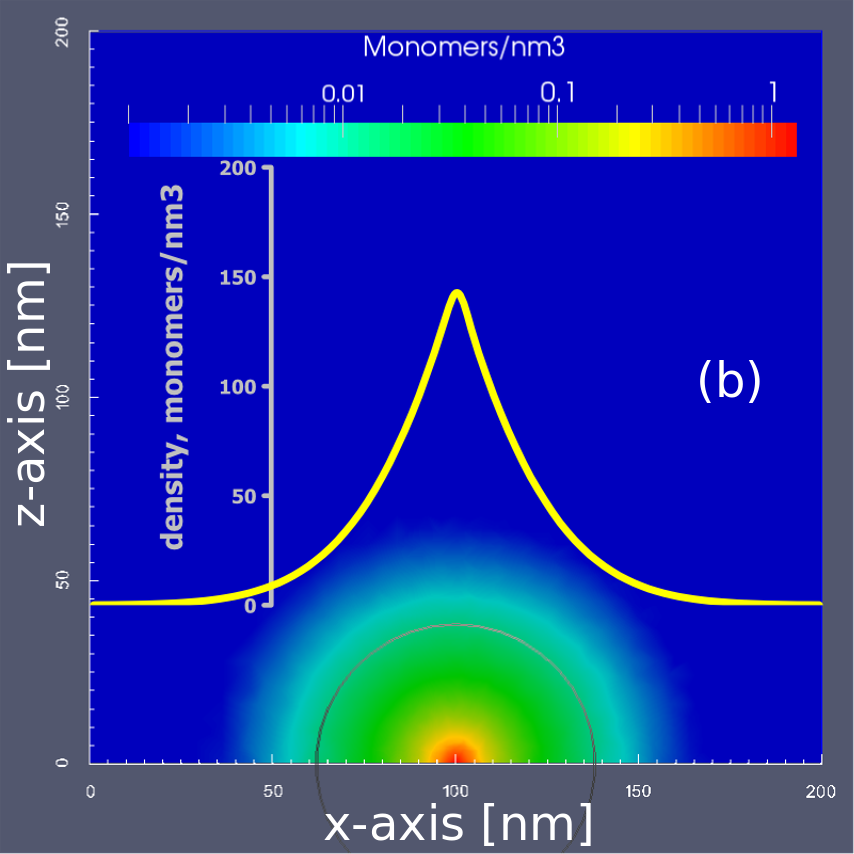} 
  \end{subfigure} 
  \begin{subfigure}[b]{0.5\linewidth}
    \centering
    \includegraphics[width=\linewidth]{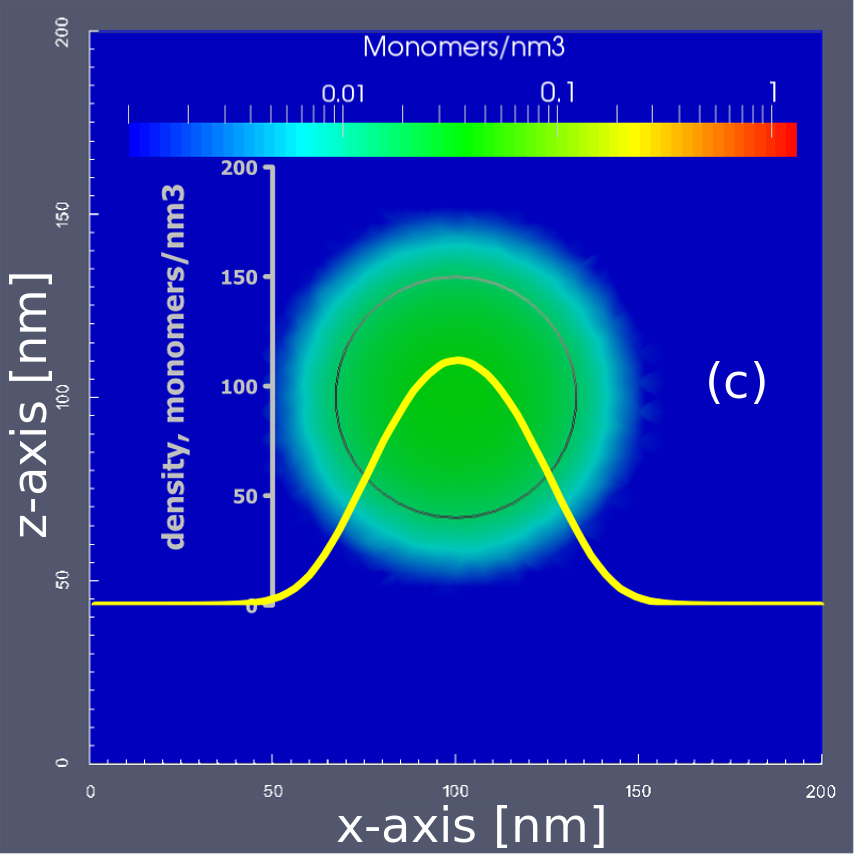} 
  \end{subfigure}
  \begin{subfigure}[b]{0.5\linewidth}
    \centering
    \includegraphics[width=\linewidth]{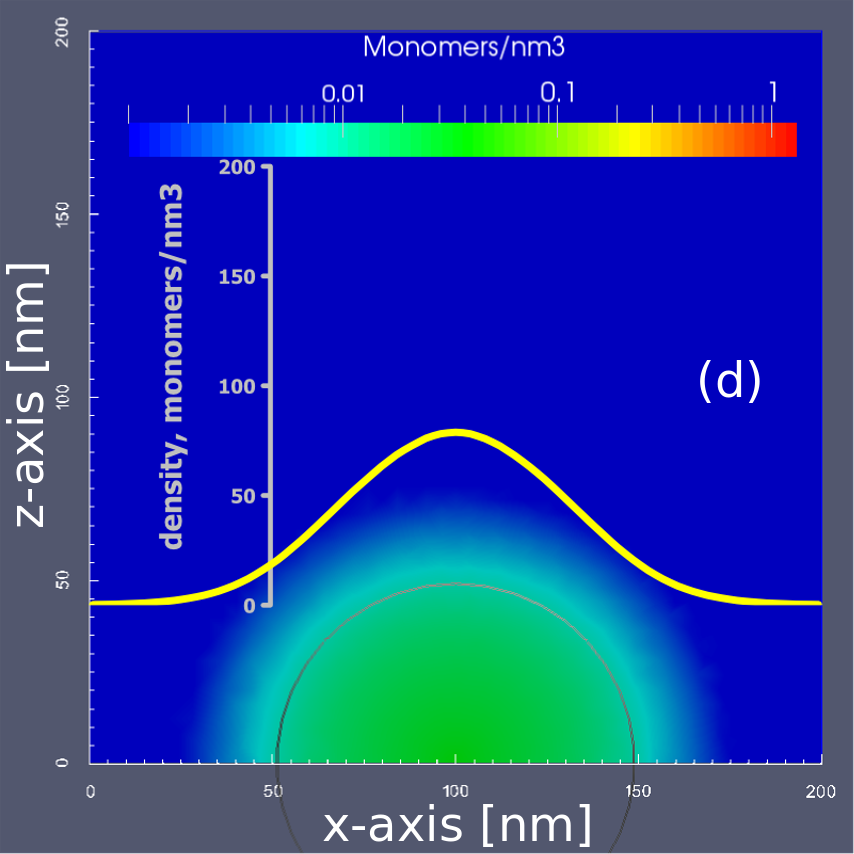} 
  \end{subfigure} 
  \caption{The effect of the excluded volume on free (panels (a,c), chain centered at (100,100,100)~nm) and grafted (panels (b,d), chain grafted at (100,100,0)~nm) chains in theta (a,b) and good (c,d) solvents. The color code shows the 2D cross-sections of the density profile, $\rho(x,100,z)$. The silver circle is the radius of gyration in each case. The thick yellow curve is the integrated density along the $x$-axis: $\rho(x) = \int dz\, dy\, \rho(x,y,z)$, displayed as an overlapping inset.}
  \label{slice} 
\end{figure}

\subsection{Monomer-substrate repulsion}
While our mushroom is electrically neutral, it may still have a short-ranged Van der Waals interaction with the substrate~\cite{de1981polymer}. One can assume that the polymer affinity to a good solvent is greater than its affinity to the wall, resulting in an effective monomer-substrate repulsion. We describe it with a potential suggested by Hamaker (point-to-plane interaction): $w(z) = H/z^3$, where $H$ is a material-dependent Hamaker constant. The potential is added to the diffusion-advection equation:
\begin{equation}\label{fulleq}
\frac{\partial q}{\partial s} = R_g^2 \nabla^2 q - \tau_r \mathbf{u}\cdot \nabla q + (vN\rho) q + \frac{H}{z^3}q, 
\end{equation}
and similarly for $q^{\dagger}$. A polymer depletion layer shows up near the substrate as anticipated. Similar profiles can also be obtained by considering other fast-decaying potentials, such as an exponential $w(z) = H/z_0^3 e^{-z/z_0}$.

We solve Eq.~\eqref{fulleq} by the standard method of finite elements (using FreeFEM++ software~\cite{hecht2012}), with the backwards-Euler marching scheme for the $s$ integral. The initial Dirac delta condition $q(\mathbf{r},s=0) = \delta(\mathbf{r})$ is numerically approximated by a very narrow Gaussian function. We have verified that the final result does not depend on the chosen width of the initial Gaussian.

\section{Results and discussion}
\begin{figure}[htb]
    \centering
    \includegraphics[width=0.8\linewidth]{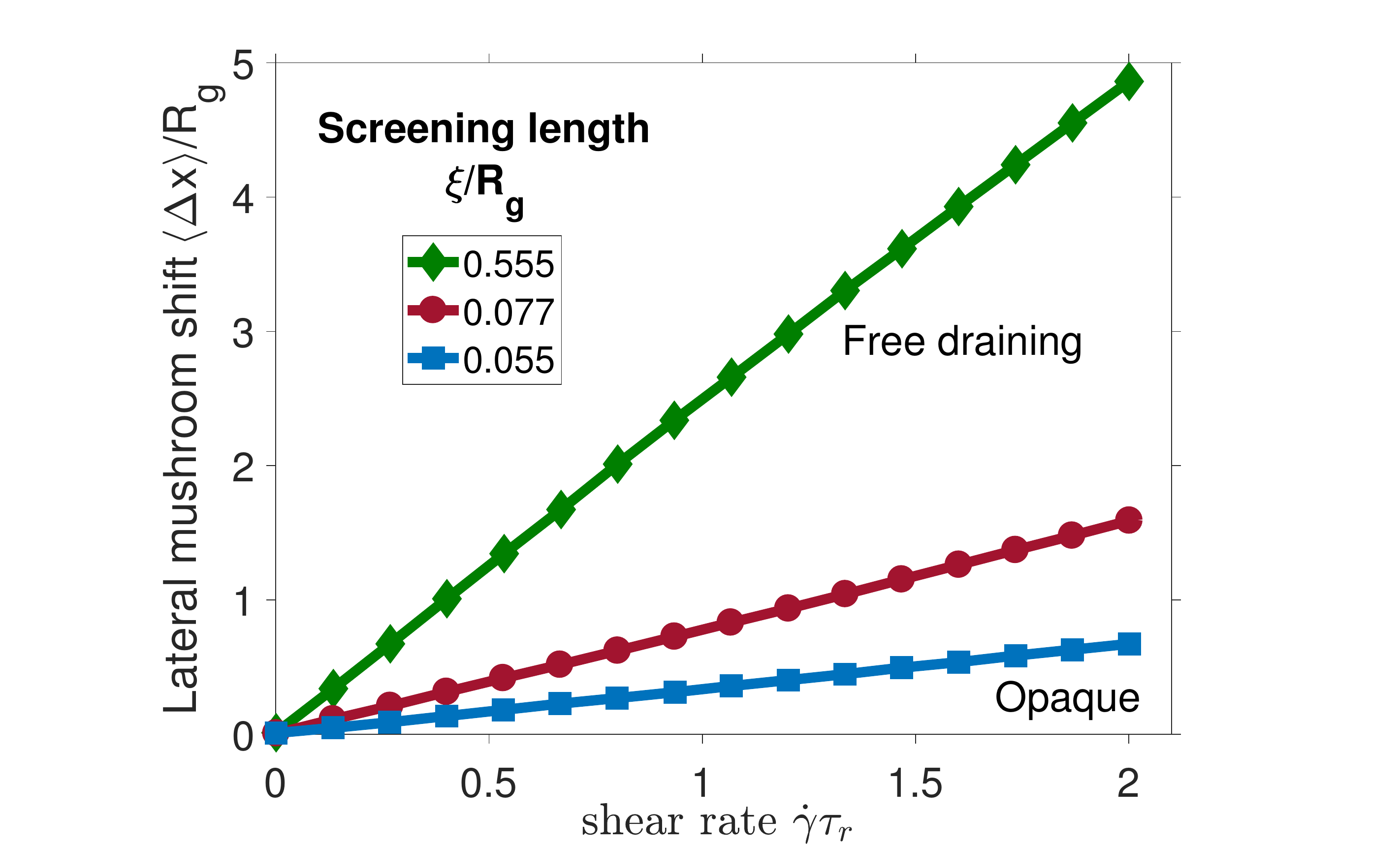}
    \caption{The lateral shift of the chain center of mass $\braket{x} = \int x\rho\, dy\, dz$, as a function of the applied shear rate normalized to the Rouse relaxation time $\dot{\gamma} \tau_r$. The shift is maximum for a free draining chain and decreases with decreasing liquid screening length $\xi/R_g$.}\label{xdev}
\end{figure}

The complete self-consistent solution with hydrodynamics, surface repulsion and excluded volume taken into account is visualized in Fig.~\ref{algorithm}. Concerning shear flow, the main conclusion is that the mushroom is mostly stretched along the direction of the flow, see Fig.~\ref{xdev}. The effect is most pronounced for high shear rate $\dot{\gamma} \tau_r \gg 1$, and low screening length $\xi/R_g \ll 1$. We could identify only one mechanism by which the flow in the $x$ direction could possibly cause any change in the $z$ density profile. When the hydrodynamic screening length $\xi/R_g$ is reduced (see Eq.~\eqref{darcy}), the fluid streamlines deviate around the chain and gain a positive $z$ velocity component which on the incoming side swells the polymer away from the substrate. On the opposing side there is a negative $z$ velocity component which compresses the polymer, resulting in an overall irregular shape.

\begin{figure}[htb]
    \centering
    \includegraphics[width=0.8\linewidth]{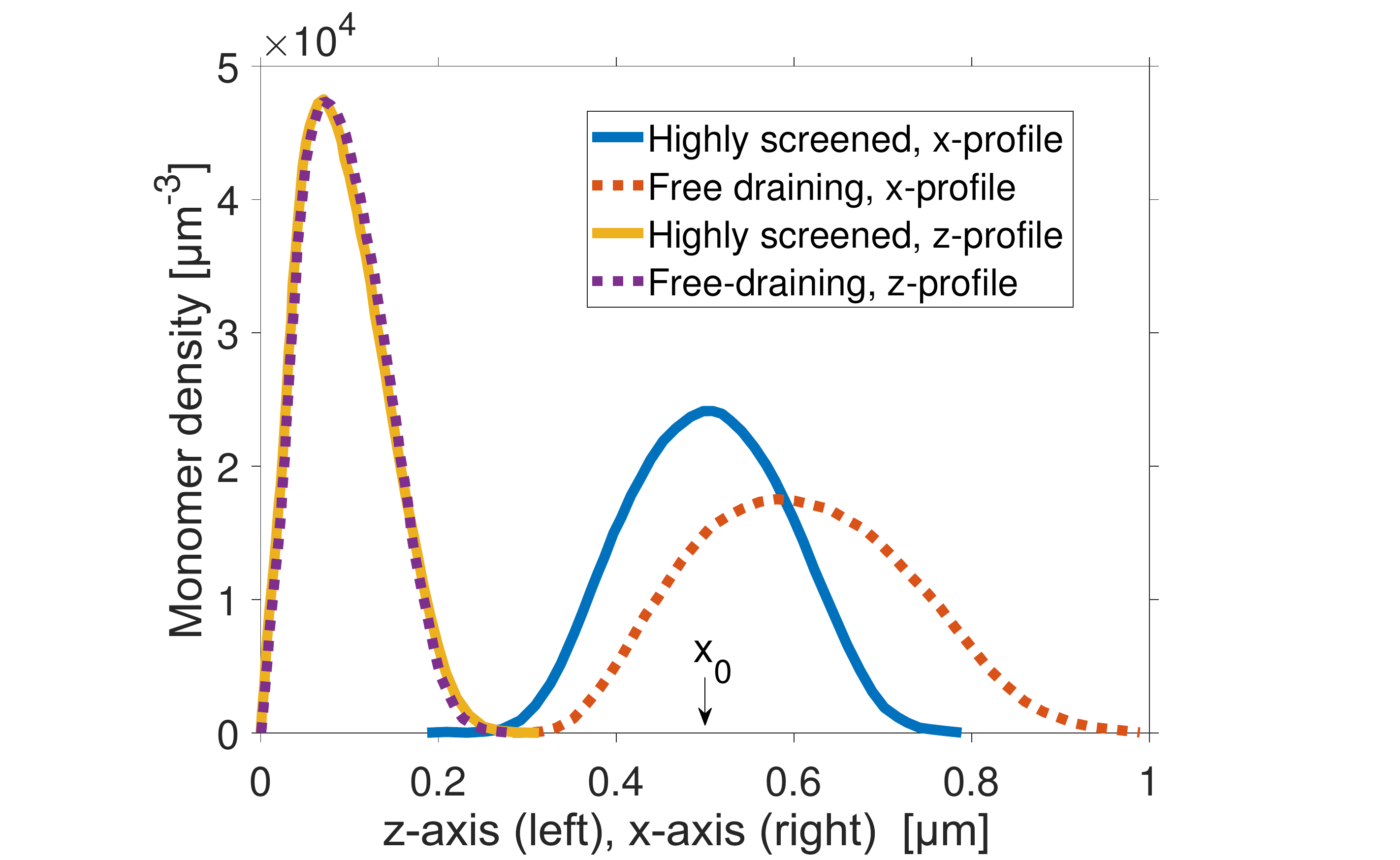}
    \caption{The polymer density profile along the flow $x$ (blue and red curves), and perpendicular to flow $z$ (yellow and purple curves). We compare a free draining polymer (dotted lines, liquid penetration length $\xi/R_g \approx 1$) against one which strongly screens the flow (solid lines, $\xi/R_g \ll 1$), under the same applied shear rate. The effect of flow is entirely along the $x$-axis, where the density profile stretches and the center of mass moves away from the grafting point $x_0=\SI{0.5}{\micro\meter}$.}\label{profiles}
\end{figure}

To compare the effect of different simulation parameters we integrate the 3D density profile over a plane to obtain the density along the $z$-axis: $\phi (z) = \iint dx\, dy\, \phi(x,y,z)$. The result is shown in Fig.~\ref{profiles}, quantifying the change in polymer shape as a function of the screening length.

Under no reasonable parameters could our mean-field model display any change in brush thickness, in agreement with NR data. In essence, we cannot find any mechanism to generate a net vertical force from a viscous solvent flow. It is true that the solvent produces some upwards drag when it strikes the front of the mushroom, but then it flows over the top, and descends back on the trailing edge, summing to a net of zero force, hence zero change in the average density profile perpendicular to flow. The only possible effect is a slight increase of ``surface roughness'', but it is negligible compared to the overall roughness of the mushroom, and cannot be expected to influence the NR signal. 

While some simulations in the literature also predict a null effect (i.e. off-lattice Monte Carlo~\cite{miao1996}), others such as self-consistent Brownian dynamics, show a decrease of brush thickness under shear~\cite{saphiannikova2000self}. A crucial difference between these models, is that the ones which show a decrease invariably include non-extensible or otherwise stiff molecular bonds, so that the polymer bends over in the flow like a rigid rod~\cite{alvarado2017nonlinear}. This effect is relevant to truly stiff molecules like carbon nanotubes or nanowires. In the case of flexible polymers like polystyrene, the bond stiffness would only come into play for shear rates comparable to the fluctuation rate of the persistence or Kuhn length, $\SI{e9}{\per\second}$ or more. Yet, the highest shear rate confirmed on Earth is \SI{e8}{\per\second}, caused by an asteroid impact~\cite{ramesh2008high}, so a flexible polymer brush will be destroyed well before it could shrink perpendicular to a Newtonian shear flow.

\section{Acknowledgements}
The authors thank Richard Michel for his guidance on the use of FreeFEM++ software. In addition, Mark Johnson and Luca Marradi have contributed by kindly providing access to computational resources. NR was conducted at Liquids Reflectometer (SNS, Oak Ridge, USA). We acknowledge the help of Candice E. Halbert, and the advice of Fr{\'e}d{\'e}ric Restagno and Liliane L{\'e}ger.

\section{Appendix: End-tethered vs. free chain in a good solvent}\label{exvolconst}
Numerous measurements of polystyrene radius of gyration in various solvents have been carried out over the years. Using the data from \citeauthor{fetters1994}\cite{fetters1994}, we can interpolate that a model chain of $M_w = \SI{6.0e5}{\gram\per\mole}$ molecular mass should have the radius of gyration in a theta solvent (cyclohexane) of $R_g^{\Theta} = \SI{22}{\nano\meter}$, while in a good solvent (toluene) it becomes  $R_g^{\text{TOL}} = \SI{33}{\nano\meter}$. This information will serve us to determine the excluded value parameter of our 3D model. Note that these data have been measured for free chains in dilute solutions, while we are primarily interested in end-tethered chains. If the $s=0$ end of an ideal Gaussian chain is constrained to the origin, the random walk statistics predict that the probability density of the $s$-th segment is a Gaussian:
\begin{equation}
\rho_{\text{tethered}}(\mathbf{r},s) = \frac{1}{(4\pi R_g^2 s)^{3/2}} \exp \left( -\frac{\mathbf{r}^2}{4R_g^2 s}\right).
\end{equation}
The density of the entire chain is the sum of its individual segment densities, normalized to the total number of segments $N = M_w/M_1 = \num{5761}$, where $M = \SI{104.15}{\gram\per\mole}$ is the molecular mass of one styrene monomer.
\begin{equation}\label{ints}
\rho_{\text{tethered}}(\mathbf{r}) = N \int_0^1 ds\, \rho_{\text{tethered}}(\mathbf{r},s).
\end{equation}
One may check that the radius of gyration of an end-tethered chain is $\sqrt{\braket{\mathbf{r}^2}} = \sqrt{3}R_g$, higher than it would be if the chain was not tethered.

The theoretical density distribution of a chain whose both ends are free is less well known, so we report it here. A specific chain conformation can in principle be described by a certain parametric function $\mathbf{r}(s)$. One can write a Fourier transform of this curve to obtain the so-called Rouse decomposition:
\begin{equation}\label{rouse}
\mathbf{r}(s) = \mathbf{a_0} + 2 \sum_{p=1}^{\infty} \mathbf{a_p} \cos (p\pi s).
\end{equation}
It satisfies the physical requirement that $(d\mathbf{r}/ds) \lvert_{s=0,s=1} \equiv 0$, that is, there can be no tension at either chain end, meaning that both ends are free to move. The vector $\mathbf{a_0}$ denotes the center of mass of the chain, while the other vectors $\mathbf{a_p}$ are all independent Gaussian random variables with the mean equal to zero and the variance equal to
\begin{equation}
\braket{\mathbf{a_p^2}} = \frac{R_g^2}{\pi^2 p^2}
\end{equation}
(see related discussion in \citeauthor{doi1986}\cite{doi1986}). According to Eq.~\eqref{rouse}, $\mathbf{r}(s)$  is a sum of Gaussian random variables, hence it itself is also a Gaussian random variable with the mean equal to $\braket{\mathbf{r}(s)} = \mathbf{a_0}$ (position of the center of mass) and the variance equal to
\begin{equation}
\braket{(\mathbf{r}(s) -\mathbf{a_0})^2} = \sum_{p,q} \braket{\mathbf{a_p \cdot a_q}} \cos(p\pi s) \cos(q\pi s) = \frac{R_g^2}{\pi^2} \sum_{p=1}^{\infty} \frac{\cos^2 (p\pi s)}{p^2} = 6R_g^2 \left(s^2 - s + \frac{1}{3} \right).
\end{equation}
Clearly, the end monomers $s=0$ and $s=1$ have a wider distribution ($\sigma_r = \sqrt{2}R_g$) with respect to the middle ones ($s=\frac{1}{2}$) which are more concentrated in the center ($\sigma_r = R_g/\sqrt{2}$). The full distribution function is hence given by
\begin{equation}
\rho_{\text{free}}(\mathbf{r},s) = \left[\frac{1}{4\pi R_g^2 (s^2-s+1/3)}\right]^{-3/2} \exp \left( -\frac{\mathbf{r}^2}{4R_g^2 (s^2-s+1/3)}\right).
\end{equation}
To obtain the distribution of the whole chain, simply integrate over $ds$, just like in Eq.~\eqref{ints}:
\begin{equation}\label{freedist}
\rho_{\text{free}}(\mathbf{r}) = N \int_0^1 ds\, \rho_{\text{free}}(\mathbf{r},s).
\end{equation}
One may check that the radius of gyration of this distribution is $\sqrt{\braket{\mathbf{r}^2}} \equiv R_g$, as expected for a free chain.

We now apply the excluded volume algorithm to the density given in Eq.~\eqref{freedist}, as described in the main text. The excluded volume parameter is progressively increased until the radius of gyration becomes equal to the experimental value $R_g^{\text{TOL}} = \SI{33}{\nano\meter}$ (the excluded volume turns out to be $v=\SI{0.016}{\cubic\nano\meter}$ per monomer). We can now use this value and apply the excluded volume algorithm to the end-tethered chain [Eq.~\eqref{ints}] which then swells up to a radius of gyration of \SI{49}{\nano\meter}. The density plots of all the polymers are shown in Fig.~\ref{slice}.

\subsection{A Gaussian chain under a uniform flow: analytical solution}\label{uniflow}
One way to illustrate the theory of polymer chains under liquid flow is to find an analytical solution for the simplest non-trivial case. Consider an ideal Gaussian chain in one dimension, whose one end is tethered at $x=0$ and the other end is free to be anywhere on the $x$-axis. A uniform flow of magnitude $u$ is applied in the $x$ direction. According to Eqs.~\eqref{adv1}-\eqref{adv2}, the partition function of such a chain satisfies the diffusion-advection (D-A) equation:
\begin{align}
\frac{\partial q}{\partial s} &= \frac{\partial^2 q}{\partial x^2} - u \frac{\partial q}{\partial x}, \qquad q(x,s=0) = \delta(x),\\
\frac{\partial q^{\dagger}}{\partial s} &= -\frac{\partial^2 q^{\dagger}}{\partial x^2} - u \frac{\partial q^{\dagger}}{\partial x}, \qquad q^{\dagger}(x,s=1) = 1,
\end{align}
written in dimensionless units for simplicity. Both of these equations have simple analytical solutions:
\begin{align}
q(x,s) &= \frac{1}{\sqrt{4\pi s}} \exp \left(-\frac{(x-us)^2}{4s}\right),\\
q^{\dagger}(x,s) &= 1.
\end{align}
To obtain the density of the entire chain, we simply integrate over each monomer:
\begin{equation}\label{DA}
\rho_{\text{D-A}}(x) = \int_0^1 ds\, qq^{\dagger} = \int_0^1 \frac{ds}{\sqrt{4\pi s}} \exp \left(-\frac{(x-us)^2}{4s}\right).
\end{equation}
On the other hand, for the simple case of a 1-D uniform flow, one might be tempted to use the modified diffusion equation [Eqs.~\eqref{mde1}-\eqref{mde2}] with a scalar potential: $w(x) = -gx$, where the parameter $g$ denotes a constant force. This approach would be correct if the force was conservative (i.e. gravity, Couloumb force), but it is invalid for a dissipative force such as the Stokes which we are dealing with in this study. Nevertheless, it is interesting to compare the MDE prediction with that of D-A. Consider now the partition function of a Gaussian chain in a uniform scalar potential:
\begin{align}
\frac{\partial q}{\partial s} &= \frac{\partial^2 q}{\partial x^2} - gxq, \qquad q(x,s=0) = \delta(x),\\
\frac{\partial q^{\dagger}}{\partial s} &= -\frac{\partial^2 q^{\dagger}}{\partial x^2} + gxq, \qquad q^{\dagger}(x,s=1) = 1.
\end{align}
These equations might remind some readers of the Schr\"{o}dinger equation in a uniform field. Luckily, they both have analytical solutions:
\begin{align}
q(x,s) &= \frac{1}{\sqrt{4\pi s}} \exp \left(-\frac{x^2}{4s} + \frac{gxs}{2} + \frac{g^2 s^3}{12} \right),\\
q^{\dagger}(x,s) &= \exp \left(gx(1-s) + \frac{g^2}{3}(1-s)^3\right).
\end{align}
Integration over $x$ yields the total partition function (which is, as required, independent of $s$): 
\begin{equation}
Q = \int_{-\infty}^{\infty} dx\, qq^{\dagger} = \exp \left(\frac{g^2}{3}\right).
\end{equation}
The density is given by the integral over all monomers, normalized to the total partition function:
\begin{equation}
\rho_{\text{MDE}}(x) = \frac{1}{Q} \int_0^1 ds\, qq^{\dagger} = \int_0^1 \frac{ds}{\sqrt{4\pi s}} \exp \left(-\frac{(x-gs(2-s))^2}{4s}\right).
\end{equation}
This equation is surprisingly similar in form to the D-A density, Eq.~\eqref{DA}. To compare them, let us find the mean position of each density distribution:
\begin{equation}
\braket{x} = \int_{-\infty}^{\infty} x \rho(x)\, dx \quad \Rightarrow \quad \braket{x}_{\text{D-A}} = \frac{1}{2}u, \quad \braket{x}_{\text{MDE}} = \frac{2}{3}g.
\end{equation}

We have chosen $u=3$ and $g=9/4$ which give the same mean $\braket{x}=3/2$ for both distributions and plotted them in Figure~\ref{uflow}. Quite remarkably, completely different physics modeled by different approaches provide very similar density profiles.

Finally, we give the expression for an unperturbed end-tethered Gaussian chain density (where both models agree if we set $g=u=0$):
\begin{equation}
\begin{split}
\rho_0 (x) &= \int_0^1 ds\, \frac{1}{\sqrt{4\pi s}} \exp \left(-\frac{x^2}{4s}\right)\\ &= \sqrt{\frac{x^2}{4}} \left[\text{erf} \left( \sqrt{\frac{x^2}{4}} - 1\right) \right] + \frac{1}{\sqrt{\pi}} \exp \left( -\frac{x^2}{4}\right).
\end{split}
\end{equation}
This function is shown as a dashed line in Figure~\ref{uflow}.

\bibliography{brush}

\end{document}